\documentclass[10pt,letterpaper,twocolumn]{article} 

\usepackage{ol2}
\usepackage[draft]{hyperref}
\usepackage{amsmath}

\usepackage{amssymb}
\usepackage{bm}
\usepackage{upgreek}
\usepackage{graphicx}

\newcommand{\cE}{\mathcal{E}}
\newcommand{\bc}[1]{\bm{\mathcal{#1}}}
\renewcommand{\c}[1]{\mathcal{#1}}

\newcommand{\be}{\begin{equation}}
\newcommand{\ee}{\end{equation}}

\begin{document}

\twocolumn[

\title{Electrostrictive counter-force on fluid microdroplet in short laser pulse}

\author{Simen {\AA}. Ellingsen and Iver Brevik}

\address{
  Department of Energy and Process Engineering, Norwegian University of Science and Technology, N-7491 Trondheim, Norway
}

\begin{abstract}
When a micrometer-sized fluid droplet is illuminated by a laser pulse, there is a fundamental distinction between two cases. If the pulse is short in comparison with the transit time for sound across the droplet, the disruptive optical Abraham-Minkowski radiation force  is countered by electrostriction and the net stress is compressive. In contrast, if the pulse is long on this scale, electrostriction is cancelled by elastic pressure and the surviving term of the electromagnetic force, the Abraham-Minkowski force, is disruptive and deforms the droplet. Ultrashort laser pulses are routinely used in modern experiments, and impressive progress has moreover been made on laser manipulation of liquid surfaces in recent times, making a theory for combining the two pertinent. We analyze the electrostrictive contribution analytically and numerically for a spherical droplet.
\end{abstract}

]


Consider a laser pulse impinging on a microdroplet of a homogeneous fluid, whose radius is $a$ and of density $\rho$, situated at the origin. Assume the pulse to be short (defined below), but long enough to be treated as a plane monochromatic wave. 
The refractive index of the droplet is $n$, its permeability $\mu=1$, and we assume it is surrounded by vacuum for simplicity. The fluid is then acted upon by two different kinds of optical forces; a volume force in the interior due to \emph{electrostriction} (ES) acting towards regions of higher electromagnetic energy density, and a surface force acting on the surface $r=a$ due to the difference in permittivity between fluid and vacuum. 

When the pulse length is sufficiently long, the ES volume force is compensated by mechanical fluid pressure, and the fluid motion becomes entirely dictated by the surface force and fluid mechanical equations of motion. This was the case in the classical experiment by Zhang and Chang\cite{zhang88}, and to our knowledge only this case has been considered in the theoretical literature\cite{lai89,brevik99,chraibi08,ellingsen12}. The time it takes for the electrostrictive compressive force to be compensated depends on the compressibility of the fluid; it is approximately the time it takes for a sound wave to traverse the droplet, $\tau_c=2a/u$ where $u$ is the speed of sound.
 
The effect has become accentuated in recent years because of the frequent use  of short laser pulses. Moreover, in some modern  applications one works with very small surface tensions, typically in a two-fluid system 
such as used by 
Delville's group, c.f.~\cite{wunenburger11} and references therein, in which surface tension is reduced to a millionth that of an air-water surface. The surface is thus sensitive to even small 
stresses.

The electromagnetic force density in a fluid is\cite{brevik79},
\begin{equation}
  {\bf f}=-\frac{1}{2}\epsilon_0\langle\cE^2\rangle\nabla \epsilon +\frac{1}{2}\left[ \langle\cE^2\rangle\rho \left(\frac{\partial \epsilon}{\partial \rho}\right)_T\right] \label{1}
\end{equation}
where $\langle\cdots\rangle$ denotes average over an optical period. According to our conventions  the constitutive relations are 
$\bc{D}=\epsilon_0\epsilon \bc{E}$, $\bc{B}=\mu_0\bc{H}$; thus  $\epsilon$ is  nondimensional, and we assume it to be real and positive. We neglect an additional Abraham term (whose existence is subject to debate), which is much smaller than the other in all typical parameter ranges \cite{brevik79}. Calligraphic typeface denotes real field quantities, so $\cE = \mathrm{Re}\{E e^{i\omega t}\}$, etc. Note $\langle \cE^2 \rangle={\textstyle \frac1{2}}|E|^2$ and so on.

Consider first the second term in Eq.~(\ref{1}),  the electrostriction term, called ${\bf f}^{\rm ES}$. We draw on the Lorentz-Lorenz relation 
to evaluate the derivative of the permittivity, whereby
\begin{equation}
  {\bf f}^{\rm ES}=\frac{1}{6}\epsilon_0{\bf \nabla}[\langle\cE^2\rangle(n^2-1)(n^2+2)]. \label{2}
\end{equation}
One sees that the effect from this force on the droplet is compressive, as it points towards higher optical intensity. The striction force, when taken alone, accelerates the fluid inward until an elastic counter pressure is established, on a time scale $\tau_c$. We recently considered pressure waves set in motion by a laser beam in a homogeneous fluid \cite{ellingsen11b}.

Secondly, there is the first term in Eq.~(\ref{1}) which may be called the Abraham-Minkowski (AM) term ${\bf f}^{\rm AM}$, as it is common for the Abraham and Minkowski energy-momentum tensors \cite{brevik79}. This term is the only one to survive after the elastic response time as discussed above. Thus it is sufficient to describe the laser-induced surface deformations observed in quasistatic or long pulse experiments, f.ex.\ Refs.~\cite{zhang88,wunenburger11,ashkin73}.

By integrating the force density (\ref{2}) across the boundary we obtain the ES surface pressure ${\bf P}^{\rm ES}$,
\begin{equation}
  {\bf P}^{\rm ES}=
  -\frac{1}{6}\epsilon_0(n^2-1)(n^2+2)\langle\cE^2(a^-)\rangle\hat{\mathbf{n}}
  \equiv\sigma^{\rm es}\hat{\mathbf{n}}, \label{4}
\end{equation}
where $\hat{ \bf n}$ is the outward normal and $\sigma^{\rm ES}$ the scalar pressure, a negative quantity. In the following we shall assume circularly polarized plane wave illumination, drawing on the Mie scattering formulation of Barton \emph{et al}.\ \cite{barton89}; the circularly polarized fields are given in full in \cite{ellingsen12}.

According to Eq.~(\ref{2}) the ES force also has a volume contribution from the interior of the sphere proportional to ${\bf \nabla} \cE^2$. One could calculate this force in detail, yet for our purposes it is physically more instructive to consider some overall properties of the ES volume force which allows simple comparison to the AM surface force.

First, note that the \emph{net} ES force on the whole sphere is zero. This is easily seen: the force \eqref{2} is a gradient and when integrated over any volume $\c{V}$ can be written as an integral over surface area. For any control volume outside the sphere, $\partial \epsilon/\partial\rho$ is zero, hence zero total force. 

Secondly, a physically instructive quantity to calculate is the \emph{full} ES force (surface and volume contributions) acting on the front and back hemispheres. Choosing a control volume bisecting the sphere at $z=0$ and closed outside the sphere for $z<0$, the force may be written as an integral over the circular section only,
\begin{equation}
  F_{z,<}^{\rm ES}=\frac{\pi}{3} \epsilon_0(n^2-1)(n^2+2)\left.\int_0^ardr\langle\cE^2(r)\rangle\right|_{\theta=\frac{\pi}{2}}, \label{13}
\end{equation}
$r$ being the radius in cylindrical coordinates. $<(>)$ denotes front (back) as seen by light propagating from $z=-\infty$. Exactly the same argument gives $F_{z,>}^{\rm ES}=-F_{z,<}^{\rm ES}$, which accords with zero total force.

Consider next the AM force ${\bf f}^{\rm AM}$, which according to Eq.~(\ref{1}) acts in the dielectric boundary layer only, where $\nabla\epsilon\neq 0$. The corresponding surface pressure ${\bf P}^{\rm AM}$ is
\begin{equation}
  {\bf P}^{\rm AM}=
  \frac{\epsilon_0}{2}(n^2-1)\langle\cE_t^2(a^-)+n^2 \cE_r^2(a^-)\rangle \hat{\mathbf{n}}
  \equiv\sigma^{\rm AM} \hat{\mathbf{n}}, \label{14}
\end{equation}
Here $\bc{E}_t=\cE_\theta \hat{\boldsymbol{\theta}}+\cE_\phi \hat{\boldsymbol{\phi}}$ is the field component parallel to the surface, and $\cE_r$ the component normal to it. The expression (\ref{14}) is positive, in accordance with the general property of the dielectric gradient force that it always acts in the direction of the optically thinner medium. 

For comparison we calculate the AM force on the front ($<$) and back ($>$) hemispheres, 
\be
  F_{z,<}^\mathrm{AM} = 2\pi a^2 \int_{\pi/2}^\pi d\theta \sin\theta \cos\theta \sigma^\mathrm{AM}(\theta)
\ee
and similarly for the back upon replacing the integral with $\int_0^{\pi/2}$. 
Unlike for ES, the two are of unequal magnitude and the net force is a push in the direction of light propagation; the sphere acts as a lens concentrating the electromagnetic energy near the rear surface. 

\begin{figure}[tb]
  \includegraphics[width=\columnwidth]{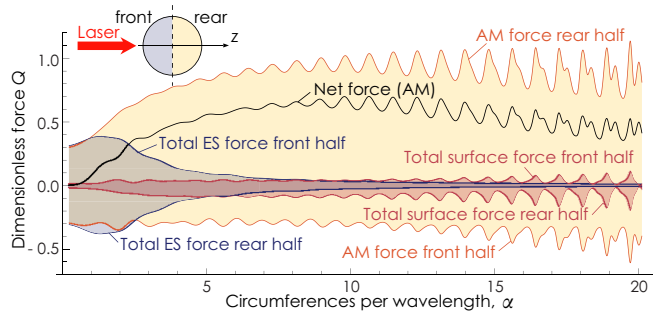}
  \caption{Force components on front and rear hemispheres. $Q$ is repulsive if positive for rear half or negative for front half.}
  \label{fig:hemispheres}
\end{figure}

The field equations for circularly polarized light are straightforward to work out in the formalism of \cite{barton89}, and are quoted, e.g., in Appendix C of \cite{ellingsen12}. 
We let $\alpha=2\pi a/\lambda_- = n\omega a/c$ be the number of circumferences per wavelength at frequency $\omega$. With field components $E_r, E_\theta, E_\phi$ written out, the front and back hemisphere AM forces can be calculated numerically, as shown in figure \ref{fig:hemispheres} where we plot dimensionless force $Q^\mathrm{AM}_{\lessgtr}=F_{z,\lessgtr}^{\rm AM}/(\pi a^2\epsilon_0E_0^2 )$, where $E_0^2$ is the rms value of the (real) electric field.

We are able to calculate the hemisphere total ES forces analytically, since at $\theta=\frac{\pi}{2}$ Legendre polynomials simplify through $P_n^1(0)=-\delta_{n1}$, $P_n^{1\prime}(0)=\delta_{n0}$ and $\psi_1(x)=x^{-1}\sin x-\cos x$. After some straightforward calculation we find $F_{z,\lessgtr}^\mathrm{ES} = \pm\pi a^2\epsilon_0E_0^2 Q^\mathrm{ES}$ with
\begin{subequations}\label{cldl}
\begin{align}
  Q^\mathrm{ES} =& 2(n^2-1)(n^2+2)\{|c_1|^2[4I_1(n\alpha)+I_2(n\alpha)]\notag \\
  &+|d_1|^2I_3(n\alpha)\}/(8\alpha^2)\\
  I_1(x) =& (2x^4-2x^2-1+\cos 2x + 2x\sin 2x)/(8x^4)\\
  I_3(x) =& {\textstyle\frac1{2}}[\gamma-1-\mathrm{Ci}(2x) + \log 2x \notag \\
  &+ x^{-2}(2x\cos x-\sin x)\sin x]\\
  I_2(x) =& I_1(x)+I_3(x) -\frac{x^2-3\sin^2x+x\sin 2x}{2x^2}.
\end{align}
\end{subequations}
Here $I_{1,2,3}$ are the radial integrals $\int_0^{n\alpha}d\varkappa$ over $\psi_1^2(\varkappa)/\varkappa^3$, $\psi_1^2(\varkappa)/\varkappa$ and $\psi_1^{\prime 2}(\varkappa)/\varkappa$, respectively, resulting from insertion of $E^2$ into Eq.~\eqref{13}. $\mathrm{Ci}$ is the cosine integral. 

\begin{figure}[tb]
  \includegraphics[width=\columnwidth]{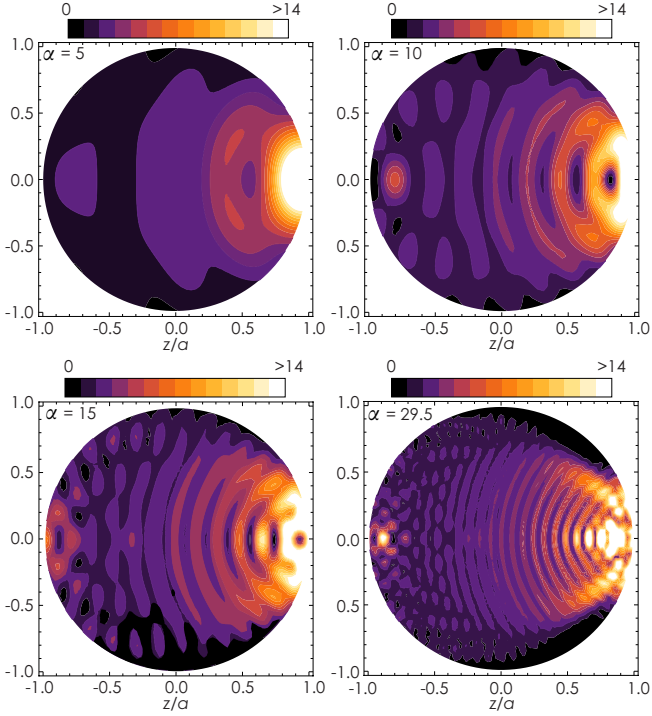}
  \caption{Electrostiction potential $\langle \cE^2\rangle/E_0^2$ inside droplet for four values of $\alpha$. Laser beam incident from left. }
  \label{fig:Esq}
\end{figure}

In figure \ref{fig:hemispheres} we compare the hemisphere forces on the front and back hemispheres due to electrostriction and electromagnetic forces. In all figures $n=1.33$. The total AM force, dictating the centre of mass motion of the droplet, is positive (pushed by the laser), whereas the total surface force 
is much smaller and of
of opposite sign, and the total surface force slightly is compressive near the focal points at the droplet's rear. ES volume forces balance the corresponding surface forces in sum since also the droplet's interior liquid is attracted to the areas of large field strength near, but not at, the droplet's rear surface. According to Eq.~\ref{2} $\langle \cE^2\rangle$ acts as a potential for the electrostrictive force density, and is shown in Fig~\eqref{fig:Esq} for four values of $\alpha$. The higher $\alpha$, the more electrostrictive compression is localised in small areas near the rear droplet boundary. 

\begin{figure}[tb]
  \includegraphics[width=\columnwidth]{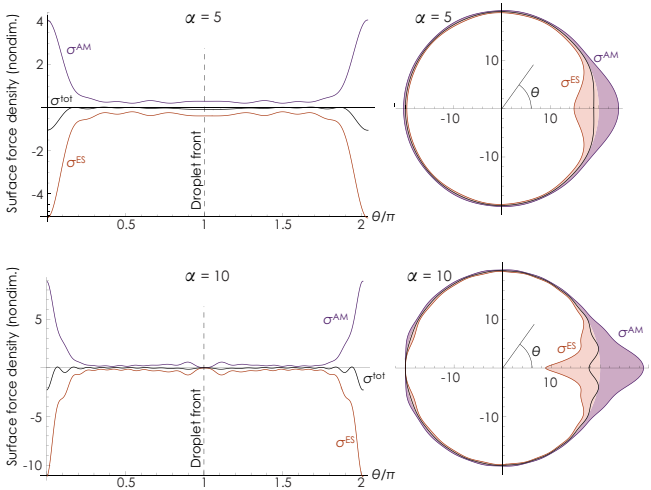}
  \caption{Surface force densities $\bar\sigma=\sigma/(\varepsilon_0 E_0^2)$ for AM and ES as functions of polar angle $\theta$. Polar plot of $r=\bar\sigma(\theta)+20$ for illustration ($20$ is an arbitrary number for visualisation).}
  \label{fig:radial}
\end{figure}

In Fig.~\ref{fig:radial} we show how optical surface forces are distributed over polar angles $\theta$, plotting $\sigma^{\rm AM}(\theta)$, $\sigma^{\rm ES}(\sigma)$ and the sum of these,  $\sigma^\mathrm{tot}$. The net surface force is compressive, which serves to 
counteract the disruptive AM force:
The AM surface force is nearly cancelled by the ES compressive force, leaving a much smaller $\sigma^\mathrm{tot}$ which is compressive on average but can take small repulsive values in certain areas. The cancellation is exact for $n\to 1$, while the net surface force density becomes more strongly compressive for increasing $n$. Note correspondence with first two panels of Fig.~\ref{fig:Esq}.

We have calculated the effect of electrostriction for the optical force on a liquid droplet. The resulting behaviour depends crucially on the laser pulse duration compared to the sonic transit time across the dimensions of the sphere. 
For example, in water at room temperature, $u\approx 1500$\,m/s, so that if $a=50$\,$\mu$m, the pulse is short in this sense when $\tau \leq 70$\,ns, which is well achievable in practice (current ultrashort pulses have femtosecond duration). For the purpose of optical pulling\cite{chen11} of microdroplets, pulses of a few ns at intervals of about one $\mu$s were found to be appropriate\cite{ellingsen12}.
Laser pulses used in droplet manipulation experiments have conventionally been longer than this, allowing the elastic counter pressure time to build, cancelling the contribution from electrostriction. Only the Abraham-Minkowski force remains, and the force is repulsive as Fig.~\ref{fig:hemispheres} shows. This is sufficient to explain experiments conducted to date, e.g.\ \cite{zhang88,wunenburger11}. At times much earlier than $t \ll 2a/u$, however, the liquid respons both to the electrostrictive and AM forces as there is no appreciable counter force met by elastic pressures (there is in this sense an analogy with the classic non-equilibrium pressure experiment of Goetz and Zahn from around 1960 \cite{goetz58}). Inclusion of electrostriction is now crucial to describing the behaviour of the droplet.

In order to describe droplet hydrodynamics at short times, we have shown that a fully compressible theory is required, whereas previous theoretical treatments were incompressible \cite{lai89,brevik99,chraibi08,ellingsen12}. What the effect on the droplet's motion from electrostriction will be is still and open question of great interest. No experiments exist to our knowledge where this has been probed. It is clear that the initial deformation is to compress the droplet. If viscosity is low, overcompensation by the elastic pressure build-up would be expected to give rise after a while to surface deformations similar to those for a longer pulse, whereas with higher viscosity we expect surface oscillations to be significantly smaller due to the conteraction of electrostriction. It seems to us that the development of a fully compressible theoretical framework for this geometry is a natural goal for the near future.

\end{document}